\documentclass{jfm}

\usepackage{amsmath}
\usepackage{bm}

\usepackage{natbib}
\usepackage{graphicx}
\usepackage{epstopdf, epsfig}
\usepackage{subcaption}
\usepackage{subcaption}
\usepackage{graphicx}
\usepackage{hyperref}
\hypersetup{
	breaklinks,
	colorlinks=true,
	linkcolor=blue,
	citecolor=purple,
	pdfauthor={Eojin Kim and Brian F Farrell},
	pdftitle={Statistical State Dynamics based study of Langmuir Turbulence}
}

\usepackage{microtype}

\usepackage{bm}

\usepackage[usenames,dvipsnames]{xcolor}

\definecolor{ao(english)}{rgb}{0.0, 0.5, 0.0}

\usepackage[normalem]{ulem}

\def\P{{\bf P}}

\newcounter{saveeqn}%

\newcommand{\be}{\begin{equation}}
\newcommand{\ee}{\end{equation}}
\newcommand{\bdm}{\begin{equation*}}
\newcommand{\edm}{\end{equation*}}
\newcommand{\bea}{\begin{eqnarray}}
\newcommand{\eea}{\end{eqnarray}}

\newcommand{\partialf}[2]
{
 \ifthenelse{\equal{#1}{}}{\frac{\partial}{\partial #2}}{\frac{\partial #1}{\partial #2}}
}

\usepackage{upgreek}
\usepackage{comment}

\renewcommand{\equiv}{\ensuremath{\stackrel{\mathrm{def}}{=}}}
\usepackage{soul}

\usepackage{subcaption}
\usepackage{graphicx}
\usepackage{pdfpages}

\begin{document}

\newtheorem{lemma}{Lemma}
\newtheorem{corollary}{Corollary}

\shorttitle{SSD based study of Langmuir turbulence} 
\shortauthor{Eojin Kim and Brian F. Farrell} 

\title{\vspace{+2ex}Statistical State Dynamics based study of Langmuir Turbulence}

\author
 {
 Eojin Kim\corresp{\email{ekim@g.harvard.edu}}\aff{1},
 Brian F. Farrell\aff{1}
  }

\affiliation
{
\aff{1}
Department of Earth and Planetary Sciences, Harvard University, Cambridge, MA~02138, USA

}

\maketitle

\begin{abstract}

The dynamics of the ocean mixed layer is of central importance in determining the fluxes of momentum, heat, gases, and particulates between the ocean and the atmosphere.  A prominent component of mixed layer dynamics is the appearance of a spanwise ordered array of streamwise oriented roll/streak structures (RSS), referred to as Langmuir circulations, that form in the presence of surface wind stress.  The coherence and long-range order of the Langmuir circulations are strongly suggestive of an underlying modal instability, and surface wind stress produces the necessary Eulerian shear to provide the required kinetic energy.  Unfortunately, there is no instability with RSS form supported solely by Eulerian surface stress-driven shear.  However, in the presence of velocity fluctuations in the water column, either in the form of a surface gravity wave velocity field and/or a background field of turbulence, there are two instabilities of the required form.  These are the Craik-Leibovich CL2 instability arising from interaction of the Eulerian shear vorticity with the Stokes drift of a surface gravity wave velocity field and the Reynolds stress (RS) torque instability arising from the organization of turbulent Reynolds stresses by a perturbing RSS.  The CL2 instability is familiar as an explanation for the RSS of the Langmuir circulation, while the RS torque instability is familiar as an explanation for the RSS in wall-bounded shear flows.  In this work, we show that these instabilities act synergistically in the mixed layer of the ocean to form a comprehensive theory for both the formation and equilibration of Langmuir circulations.

\end{abstract}

\section{Introduction}
Langmuir circulations are wind-aligned roll/streak structures (RSS) that arise in the near-surface water column in association with wind-driven shear stress and gravity waves.  In lakes and the upper ocean, the underlying structure of the Langmuir circulation is strikingly revealed at the surface by windrows of seaweed and foam. Since Langmuir's inference of this RSS from field observations (Langmuir 1938), decades of research have explored its dynamics, formation mechanism, and impact on upper-ocean mixing.

Theoretical understanding of Langmuir circulation formation is currently based on the Craik-Leibovich (CL) equation describing the mean-field dynamics of an Eulerian shear flow coexisting with a Lagrangian Stokes drift due to surface gravity waves \citep{Craik 1976, Leibovich 1980, Leibovich 1983}.   Specifically, the accepted explanation is the CL2 instability supported by the CL equation \citep{Leibovich 1983}.  

Observational evidence supports some key features of Langmuir circulations predicted by the CL2 theory. These include surface convergence and an approximate alignment of roll circulation cells with the prevailing wind.  However, observations also show Langmuir rolls penetrate more deeply into the mixed layer than roll forcing proportional to the Stokes shear CL2 theory predicts \citep{Plueddemann 1996, Smith 1996}.  
This under-prediction is exacerbated in scaling and modeling studies by using standard estimates for Stokes drift velocity ($\frac{4\pi^2 a^2}{\lambda T}$), and e-fold depth, 
($\frac{\lambda}{4\pi}$) for waves of height, $a$, wavelength $\lambda$ and period $T$,  which underestimate the surface Stokes drift and overestimate the e-folding depths for observed sea states by an order of magnitude. 
Observations give most probable values of the surface ocean Stokes drift ranging from  1 to 18 cm/s and the Stokes e-folding depth from 0.5 to 3 m \citep{Tamura 2012}.   
This surface concentration of roll forcing by the CL2 mechanism has led to the concept that the downward branch of the associated roll continues ballistically through the depth of the mixed layer without dynamic support \citep{Polton 2007}.   Regardless of whether this is a viable explanation for the Langmuir penetration discrepancy, it is clear that existence of an unstable perturbation mode with RSS form does not explain what are essentially all the observations that are advanced to support the CL2 instability, as these are of fully equilibrated finite-amplitude structures.  
However,  an unstable mode is as likely to equilibrate with a completely different structure, such as a turbulent flow, as it is to equilibrate with a structure identical to its instigating instability.  It follows that a theory for Langmuir circulations can not be comprehensive unless it predicts also the observed finite amplitude equilibria following from its predicted perturbation instabilities and the statistical structure of these equilibria; whether fixed point, limit cycle, or turbulent.  
  
While there exists a plethora of evidence for the existence of streamwise oriented rolls in the upper ocean, the roll/streak structure is the optimally growing structure in shear flow \citep{Butler-Farrell-1992} and this same structure arises in diverse physical contexts including convection \citep{Wurman 1998, Morrison 2004}, frontal regions \citep{Savelyev 2018}, and wall-bounded shear flows \citep{Farrell-Ioannou-2016-sync, Farrell-Kim-2022, Butler-Farrell-1992}.   Among alternative RSS forcing mechanisms an attractive companion to the CL2 mechanism with the potential to provide a more comprehensive theory for Langmuir circulations is the Reynolds stress (RS) torque mechanism, which was developed recently using  statistical state dynamics (SSD) methods and applied to RSS formation in turbulent shear flows \citep{Farrell-Ioannou-2016-bifur, Farrell-Ioannou-2012, Wang 2024}.

In this work, we use SSD to study the formation of  Langmuir circulations under CL dynamics.  Using SSD we are able to study the synergy between the CL2 and RS torque instabilities in the formation process and also to include in our analysis the equilibration of these instabilities to form statistical steady Langmuir equilibria  including fixed points, quasi-periodic, and turbulent states.\\

\section{Craik-Leibovich Equation}
Consider the equation of motion for an unstratified non-rotating incompressible fluid  with an imposed gravity-wave driven Stokes drift, $\underline{u_s}$:\\

 \begin{equation}
 \frac{\partial \underline{u}}{\partial t}+(\nabla \times (\underline{u}))\times (\underline{u}+\underline{u_s})=\nu \nabla^2 \underline{u}-\nabla p -\nabla(\underline{u} \cdot \underline{u_s})-\frac{1}{2}\nabla u^2
  \end{equation}\\
Where $\underline{u}=u\hat{i}+v\hat{j}+w\hat{k}$ and $\nu$ is the kinematic viscosity. Assuming the imposed Stokes drift is in the streamwise direction and varies only in the cross-stream $\underline{u_s}= U_s(y)\hat{i}$ this equation simplifies to: 
\begin{equation}\frac{\partial \underline{u}}{\partial t}+(\underline{u} \cdot \nabla )\underline{u}=\nu \nabla^2 \underline{u}-\nabla p + \begin{bmatrix}-\frac{\partial u}{\partial x}U_s \\ -\frac{\partial v}{\partial x}U_s-u\frac{d U_s}{d y} \\ -U_s \frac{\partial w}{\partial x} \end{bmatrix}\end{equation}
Assuming characteristic Eulerian velocity, $G$, and characteristic Stokes velocity, $G_{stk}$,  nondimensional parameters are the Reynolds number and the ratio of the Stokes to Eulerian velocities:\\

$$Re=\frac{GL}{\nu}$$
$$S=\frac{G_{stk}}{G},$$\\

the CL equation for our problem in dimensionless form is:\\

 \begin{equation}
 \label{eq: equationofmotion}
 \frac{\partial \underline{u}}{ \partial t}+(\underline{u}\cdot \nabla)\underline{u}=- \nabla \ p+ \frac{1}{Re}\nabla^2 \underline{u} + S \begin{bmatrix}-\frac{\partial u}{\partial x}U_s \\ -\frac{\partial v}{\partial x}U_s-u\frac{d U_s}{d y} \\ -U_s \frac{\partial w}{\partial x} \end{bmatrix}.
 \end{equation}\\

\vspace{0.25in}
\section{SSD formulation}
\vspace{0.075in}

In order to study synergy between the CL2 instability and the Reynolds stress torque instability in forming and maintaining the Langmuir RSS we now formulate the S3T SSD for the CL equation.  Taking the streamwise mean for our Reynolds average isolates the CL2 instability in the mean (first cumulant) equation while closing the expansion in cumulants at second order using a stochastic parametrization conveniently incorporates the Reynolds stress torque instability arising from interaction between the first and second cumulant.

\vspace{0.25in}
\subsection{\bf{Mean Equation}}
\vspace{0.075in}

We can express the Eulerian velocity, $\underline{u}$, in mean and perturbation form as\\
$$\underline{u}=U(y,z,t)\hat{i}+V(y,z,t)\hat{j}+W(y,z,t)\hat{k}+ \underline{u'}(x,y,z,t)$$\\
We take capital letters to denote a Reynolds averaged variable, we indicate our choice of the streamwise average for our Reynolds average so that:\\
\begin{equation}
\overline{\underline{u}}=[\underline{u}]_x=U(y,z,t)\hat{i}+V(y,z,t)\hat{j}+W(y,z,t)\hat{k}
\end{equation}
\begin{equation}
\end{equation}
We choose our domain to be periodic in the $x$ and $z$ direction, and constrained by boundaries in $y$.

Taking the $x$ average of equation \eqref{eq: equationofmotion}  allows this mean equation to be expressed as:\\

\begin{subequations}
\label{eq:mean equation}
 \begin{align}
&U_t=U_y\Psi_z-U_z\Psi_y-\partial_y(\overline{u'v'})-\partial_z(\overline{u'w'})+\Delta_1\frac{U}{Re} \\
&\Delta_1 \Psi_t=(\partial_{yy}-\partial_{zz})(\Psi_y\Psi_z- \overline{v'w'})-\partial_{yz}(\Psi_y^2-\Psi_z^2+\overline{w'^2}-\overline{v'^2})+\Delta_1\Delta_1\frac{\Psi}{Re}+S \frac{d U_s}{dy}\frac{\partial U}{\partial z}  
 \end{align} 
 \end{subequations}
where $\Psi$ is the streamfunction s.t. $\Psi_y=W$ and $-\Psi_z=V$. and 
$\Delta_1:=\partial_{yy}+\partial_{zz}$.\\ 

Boundary condition for $U,\Psi$ are
$$y=-1: U=\Psi=\frac{\partial \Psi}{\partial y}=0$$
$$y=1: U=\Psi=\frac{\partial \Psi}{\partial y}=0$$\\

\vspace{0.25in}
\subsection{\bf{Perturbation Equation}}
\vspace{0.075in}

Pressure is eliminated from the perturbation equations by exploiting nondivergence to express the equations in terms of wall normal velocity and vorticity, $v'$ and $\eta'$  
(c.f. \citep{Schmid-Henningson-2001})\\

Defining 
$$v'=\sum_k v'_k(y,z,t) e^{ikx}$$
$$\eta'= \sum_k \eta'_k(y,z,t) e^{ikx}$$\\

for perturbation state $\phi_k=[v'_k,\eta'_k]^{T}$, the perturbation equation is:\\
 
\begin{equation}\frac{\partial \phi_k}{\partial t}=A \phi_k + \epsilon^{1/2}F_k \delta\end{equation}
in which a stochastic excitation $\epsilon^{1/2}F_k \delta$ is included as a parameterization for the perturbation-perturbation nonlinearity neglected in the linearization as well as any exogenous mechanisms maintaining a field of background turbulence.  In this closure  $F_k$ is a  matrix with columns representing the spatial structure of the turbulence excitation and $\delta$ is a vector of independent temporally white noise variables.  
The scalar parameter $\epsilon^{1/2}$ is included to allow the excitation of the turbulence to be conveniently varied.\\

The operator governing the perturbation dynamics is:\\

 \begin{equation}A=\begin{bmatrix}
         L_{OS}(U)+S\cdot B_{11} &L_{C1}(U)+S \cdot B_{12} \\ L_{C2}(U)+S \cdot B_{21} & L_{SQ}(U)+S \cdot B_{22}
        \end{bmatrix},\end{equation}\\
        
        in which appear the definitions:


\begin{equation}B=\begin{bmatrix}
         B_{11} &B_{12} \\ B_{21} & B_{22}
        \end{bmatrix}\end{equation}
        \begin{equation}B_{11}=(\nabla^2)^{-1}(-U_s(ik_x) (\frac{\partial^2}{\partial x^2}+\frac{\partial^2}{\partial z^2})+ ik_x (\Delta_2^{-1})U_s(k_x^2-\frac{\partial^2}{\partial z^2})\frac{\partial^2}{\partial y^2})\end{equation}
        \begin{equation}B_{12}=(\nabla^2)^{-1}(\Delta_2^{-1} \frac{\partial U_s}{\partial y}(k_x^2-\frac{\partial^2}{\partial z^2})\frac{\partial}{\partial z})\end{equation}

\begin{equation}B_{21}=0
\end{equation}
\begin{equation}B_{22}=i k_x U_s \Delta_2^{-1}(k_x^2-\frac{\partial^2}{\partial z^2})\end{equation}\\
with $L_{OS}$ and $L_{SQ}$  the Orr-Sommerfield and Squire Operators. Expressions for   $L_{OS},L_{C1},L_{C2},L_{SQ}$ can be found in \citet{Farrell-Ioannou-2012}.  Operators $\Delta_2=\partial_{zz}-k^2$ and $\nabla^2$ are made invertible taking into account the following boundary conditions:

$$y=-1: v'=\frac{\partial v'}{\partial y}=\eta'=0$$
$$y=1: v'=\frac{\partial v'}{\partial y}=\eta'=0$$\\

Having the operator for the perturbation dynamics, $A$,  allows us to write an equation for the evolution of the perturbation correlations:\\

$$C_{k}=\phi_k \phi_k^{\dagger} $$\\   
in which $\dagger$ denotes Hermitian transpose.  
This equation has the form of the time dependent Lyapunov equation \citep{Farrell 2019}:\\

\begin{equation}
\label{eq:correlation equation}
 \frac{d C_k}{dt}=A C_k + C_k A^{\dagger}+ \epsilon Q
\end{equation}\\
with $Q \equiv F_k F_k^\dagger$.\\

Coupling the mean equation \eqref{eq:mean equation}  and the perturbation equation \eqref{eq:correlation equation}  completes the S3T SSD formulation.  This SSD has specific representation:\\

\begin{subequations}
\label{eq:S3T equations}
 \begin{align}
&U_t=U_y\Psi_z-U_z\Psi_y-\partial_y(\overline{u'v'})-\partial_z(\overline{u'w'})+\Delta_1\frac{U}{Re} \\
&\Delta_1 \Psi_t=(\partial_{yy}-\partial_{zz})(\Psi_y\Psi_z- \overline{v'w'})-\partial_{yz}(\Psi_y^2-\Psi_z^2+\overline{w'^2}-\overline{v'^2})+\Delta_1\Delta_1\frac{\Psi}{Re}+S \frac{d U_s}{dy}\frac{\partial U}{\partial z} \\
& ~~~~~~~~~~~~~~~~~~~~~~~~~~~~~~~~~ \frac{d C_k}{dt}=A C_k + C_k A^{*}+ \epsilon Q
 \end{align} 
 \end{subequations}\\

The spatial covariance of the excitation structures, $Q$, is chosen to excite each degree
of freedom with unit kinetic energy which is
accomplished by choosing Q as follows \citep{Farrell-Ioannou-2012}
\begin{equation}Q=M_k^{-1}\end{equation}
\begin{equation}M_k=(L_{u'}^{k\dagger}L_{u'}^{k}+L_{v'}^{k\dagger}L_{v'}^{k}+L_{w'}^{k\dagger}L_{w'}^{k})/(2*Ny*Nz)\end{equation}\\

 An important  observation is that the Reynolds stresses appearing in the equations for the evolution of the mean state in the first cumulant \eqref{eq:S3T equations}   can be obtained directly from the second cumulant covariance, $C_k$; for example:\\

$$\overline{u'v'}|_k=diag(L_{u'}^{k}C_k L_{v'}^{k\dagger})$$\\
in which $L_{u'}^{k}$ and $L_{v'}^{k\dagger}$ are linear differential operators.\\

The S3T SSD  \eqref{eq:S3T equations} is non-linear, but it can be straightforwardly linearized about a stationary solution so that eigenanalysis can be used to obtain SSD modes and growth rates which allows for the location of bifurcation points as a function of parameters \citep{Farrell-Ioannou-2016-bifur}.  It is important to note that the covariance being linearly perturbed is a nonlinear variable, and both the mean flow and the covariance are adjusted in the eigenanalysis, so this SSD mode is nonlinear.  This manifold of nonlinear modes  provides the basis for a comprehensive theoretical analysis of turbulence in shear flow \citep{Farrell 2025}.  
\section{S3T stability formulation}
S3T equations \eqref{eq:S3T equations} can be expressed compactly as:\\ 

\begin{equation}
\frac{d \Gamma}{dt}=G(\Gamma)+\sum_k L_{RS}C_k 
\label{compositemeaneq}
\end{equation}
\begin{equation}
\frac{d C_k}{dt}=AC_k+C_kA^{\dagger}+\epsilon Q
\label{compositecovarianceeq}
\end{equation}\\
in which the stream-wise mean state (first cumulant), $\Gamma:=[U, \Psi]^{T}$, has nonlinear dynamics:\\

\begin{equation}
G(\Gamma)=\begin{bmatrix} U_y\Psi_z-U_z\Psi_y+\Delta_1\frac{U}{Re} \\
(\partial_{yy}-\partial_{zz})(\Psi_y\Psi_z)-\partial_{yz}(\Psi_y^2-\Psi_z^2)+\Delta_1\Delta_1\frac{\Psi}{Re}+S\frac{d U_s}{dy}\frac{\partial U}{\partial z}  \end{bmatrix}
\end{equation}\\
and interacts with the fluctuation equation through the Reynolds stress term:
\begin{equation}
L_{RS}C_k= \begin{bmatrix}
-\partial_y \overline{u'v'}|_k-\partial_z\overline{uw}|_k \\ \Delta_1 ^{-1}[(\partial_{yy}-\partial_{zz})(-\overline{v'w'}|_k)-\partial_{yz}(\overline{w'^2}|_k-\overline{v'^2}|_k)]\end{bmatrix}
\end{equation}\\
in which $L_{RS}$ is the linear operator of the Reynolds stress forcing as a function of $C$.
Assuming an equilibrium S3T state, $(\Gamma_e,C_e)$, in which the LHS of  \eqref{compositemeaneq} and  \eqref{compositecovarianceeq} vanish:
\begin{equation}
\Gamma_e=\begin{bmatrix}U_e\\ \Psi_e\end{bmatrix}
\end{equation}
\begin{equation}
C_e=\sum_k C_{ke},
\end{equation}
and following \citep{Farrell-Ioannou-2012,Farrell 2019} perturbation equations linearized around the SSD equilibrium state  ($\Gamma_e$, $\sum_kC_{ke}$) can be obtained:
\begin{equation}(\delta \Gamma)_t=\sum_{i}^{} \frac{\partial G}{\partial \Gamma_i}|_{\Gamma_{e}} \delta \Gamma_i+ \sum_{k}^{}L_{RS}\delta C_k 
\label{meanS3Tlinear}
\end{equation}
\begin{equation}(\delta C_k)_t=A_{ke}\delta C_k+ \delta C_k A_{ke}^{\dagger}+ \delta A_{k} C_{ke}+C_{ke}\delta A_{k}^{\dagger}\label{covarianceS3Tlinear}\end{equation}
where 
\begin{equation}\delta A_k=A_k(\Gamma_e+ \delta \Gamma)-A_k(\Gamma_e).\end{equation}\\

Equations \eqref{meanS3Tlinear} and \eqref{covarianceS3Tlinear} comprise the formulation of the linear perturbation S3T dynamics.\\

We choose for our Langmuir model $Re=300$. Adjustable parameters are the Stokes-to-Eulerian shear ratio, S, and the intensity of  background turbulence excitation, $\epsilon$. 

There are two mean velocity profiles in this problem: the Eulerian  streamwise mean flow $U(y,z,t)$, which in the linear problem will be assumed to be a constant shear, $U(y)=y$; and the Stokes drift, which we will take to be a constant shear parameterized by that shear, $S$;  so that where the Stokes drift occurs, $u_s=Sy$. Our simulations have resolution $Ny = 41$ by
$Nz = 40$ in the wall normal and spanwise directions. For convenience, Q in equation \eqref{eq:correlation equation}
is replaced by $\hat{Q}$ such that $\epsilon = 1$ results in volume averaged RMS perturbation velocity
being $1\%$ of the maximum velocity of the Couette profile. Explicitly, 
$\sqrt{2 < E_k >}=0.01 $ where $< E_k >= trace(M_kC_k)$ represents ensemble average kinetic energy density of
the perturbation field as in \citep{Farrell-Ioannou-2012}

\section{RSS instability in the S3T SSD implementation of CL dynamics}
This section presents our findings on RSS instability in CL dynamics obtained using numerical implementation of the S3T SSD perturbation equations; \eqref{meanS3Tlinear} and \eqref{covarianceS3Tlinear}.
\begin{figure}
\subfloat[]{%
            \includegraphics[width=.48\linewidth]{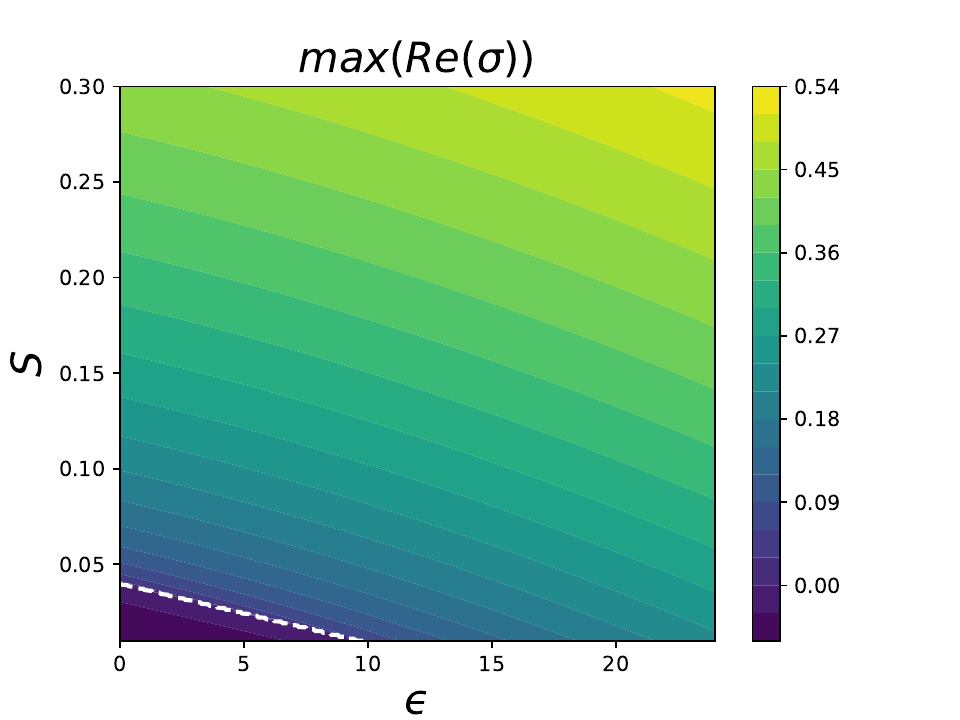}
            \label{subfig:1a}%
        }\hfill
        \subfloat[]{%
            \includegraphics[width=.48\linewidth]{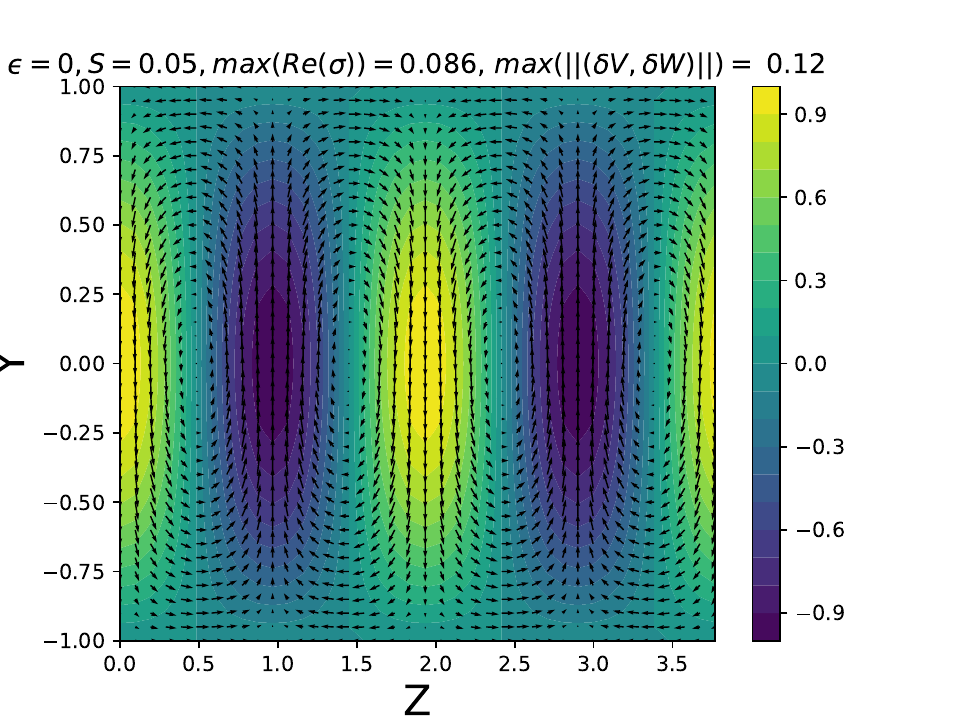}
            \label{subfig:1b}%
        }\\
        \subfloat[]{%
            \includegraphics[width=.48\linewidth]{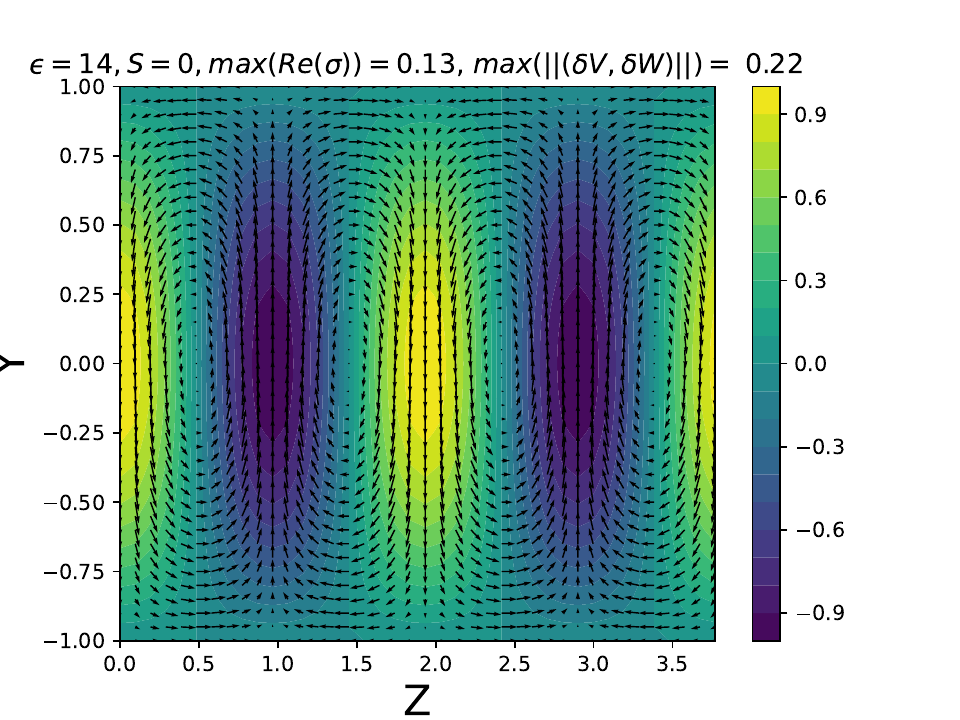}
            \label{subfig:1c}%
        }\hfill
        \subfloat[]{%
            \includegraphics[width=.48\linewidth]{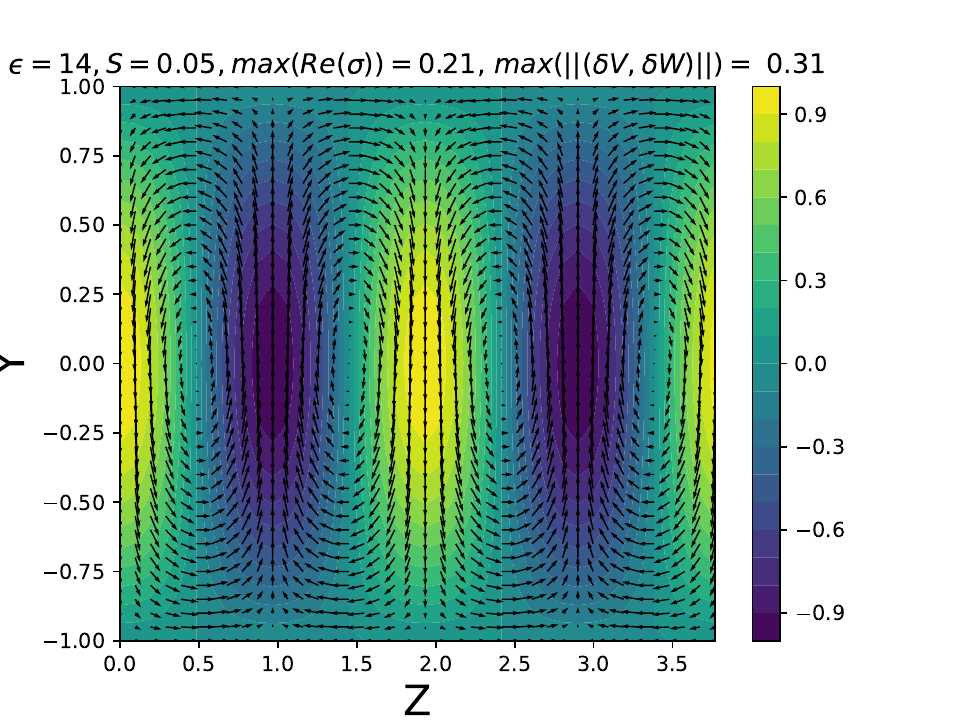}
            \label{subfig:1d}%
        }
      
\caption{Structure and stability of the synergistically interacting CL2 and RS torque instabilities. In (a) is shown the stability diagram in parameter space $(\epsilon,S)$. The RS mechanism is controlled by the background turbulence intensity parameter, $\epsilon$,  while the CL2 mechanism
is controlled by the Stokes shear, S. Contours show maximum growth rate $max(Re(\sigma))$ with $\sigma$ being the eigenvalue. The white dashed line indicates neutral stability.  Shown in $(b),(c),$ and $(d)$ are the RSS of the most unstable eigenmode at salient points in $(\epsilon, S)$. The eigenmodes have been normalized by the maximum value of the perturbation streamwise velocity $\delta U$. The perturbation streamwise velocity is shown with contours while the perturbation roll $(\delta V, \delta W)$ is shown with vectors.   In (b) is shown the RSS  supported solely by CL2 mechanism at $\epsilon=0$ and $S=0.05$. In $(c)$ is shown the RSS  supported solely by the RS torque  mechanism at $\epsilon=14$ and $S=0$. In $(d)$ is shown the RSS supported by a mixed instability at $\epsilon=14$ and $S=0.05$.}
\label{fig:Svsepsinstability}
\end{figure}

Shown in figure \ref{fig:Svsepsinstability} is the growth rate and structure of the instability with RSS form resulting from synergistic interaction between the CL2  and RS torque destabilization mechanisms. The RSS mode arising from the pure CL2 instability mechanism is shown in \ref{subfig:1b}.  The RSS mode of the pure RS torque instability mechanism is shown in \ref{subfig:1c}.  An example of an unstable RSS mode supported by synergistic interaction between the  CL2 and RS torque destabilization mechanisms is shown in \ref{subfig:1d}. It is clear from comparing these RSS modes that the  RSS arising from the CL2 mechanism, the RS torque mechanism, and their synergistic interaction are similar. 
The fact that the CL2 instability and the RS torque instability mechanisms synergistically interact to form similar RSS suggests that a comprehensive theory explaining observed Langmuir circulations can be constructed in which account is taken of the operation of both these mechanisms.  The relative importance of these mechanisms in a given observational context would depend on wind stress, surface water wave length and amplitude, background turbulence level, and any Eulerian shears not associated with wind stress.

In this section, we used linear perturbation analysis of the S3T SSD to study the instability of RSS formation in the context of CL dynamics.   In the next section, we turn our focus to the equilibration of these instabilities in the non-linear S3T SSD.

\section{RSS equilibration in the S3T SSD implementtion of CL dynamics}
\begin{figure}
\centering{
\includegraphics[width=\linewidth]{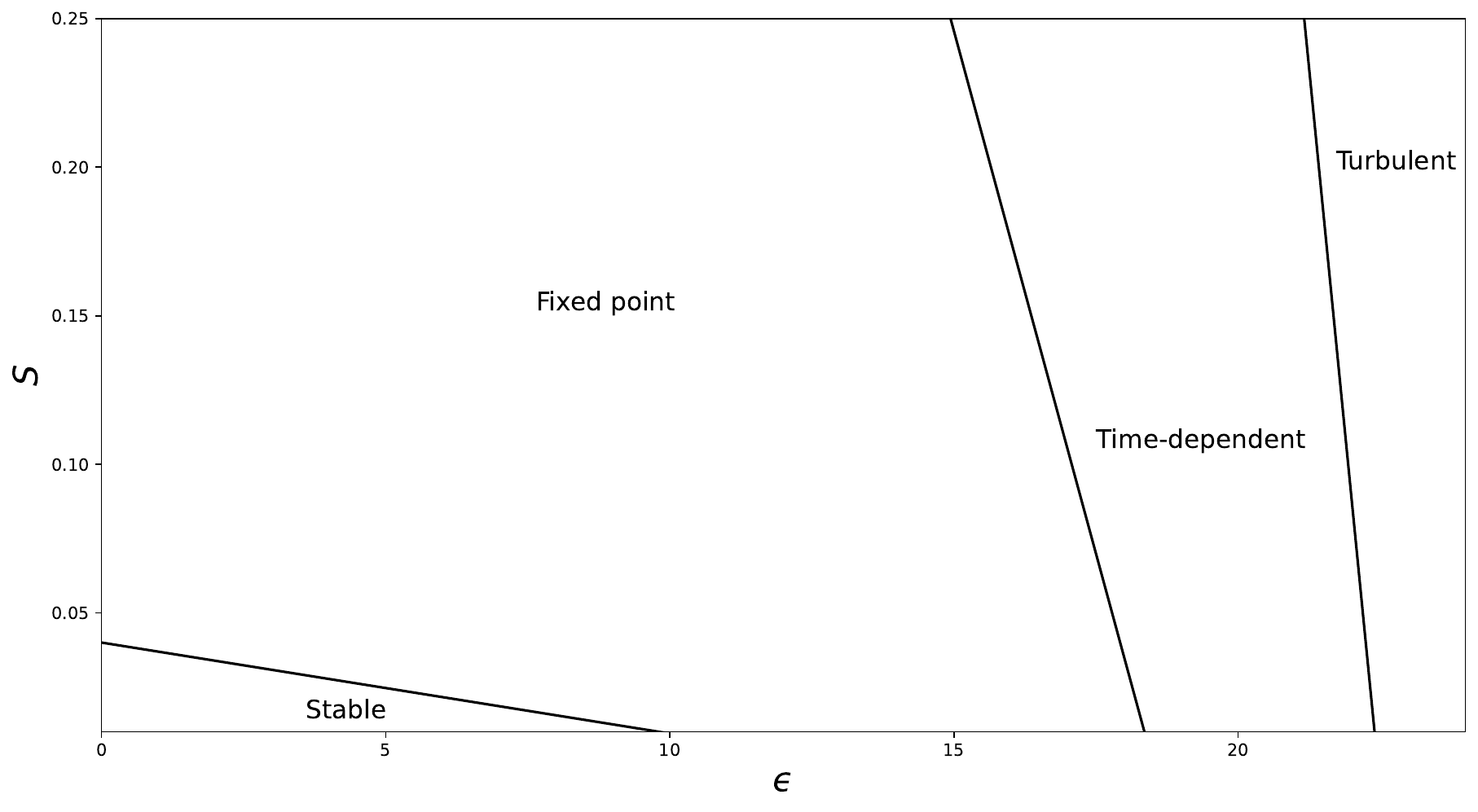}}
\caption{Regime diagram showing equilibrium RSS state as a function of background turbulence intensity parameter, $\epsilon$, and Stokes drift shear, $S$.  These parameters control the strength of the RS torque mechanism and the CL2 mechanism, respectively. At sufficiently small parameter values no RSS is supported.  With moderate $\epsilon$ and $S$, fixed-point RSS equilibria are supported. As these parameters increase a transition to time-dependent behavior is seen.  A turbulent RSS regime emerges at high parameter values; $Re=300$.}
\label{Svsepsequilibrium}
\end{figure}

At parameter values unstable to RSS, perturbing
the Langmuir turbulence SSD model results in the excitation of unstable RSS eigenmodes. These instabilities may equilibrate to fixed point RSS, to time-dependent RSS, or to turbulent
states, depending on the parameter regime. As a summary of our findings and to set the
stage for presenting our results, an equilibrium structure diagram as a function of $\epsilon$ and
S is shown in figure \ref{Svsepsequilibrium}.

We turn now to examining some dynamically salient regions of
this diagram. Indicative of finite amplitude RSS state regimes are time series of the perturbation energy:\\

\begin{equation}
TKE(t)=\frac{1}{2}[u'^2+v'^2+w'^2]_{x,y,z} 
\end{equation}\\

the streak energy:

\begin{equation}
E_{st}(t)=\frac{1}{2}[U_{st}^2]_{y,z},
\end{equation}\\

in which we have defined the streak as:

\begin{equation}
U_{st}=U-[U]_z,\\
\end{equation}\\

and the roll energy:\\

\begin{equation}
E_{r}=\frac{1}{2}{[(V^2+W^2)]_{y,z}}
\end{equation}\\

\begin{figure}
\centering{
\begin{subfigure}{0.8\textwidth} 
\centering{\caption{}\includegraphics[width=0.75\linewidth]{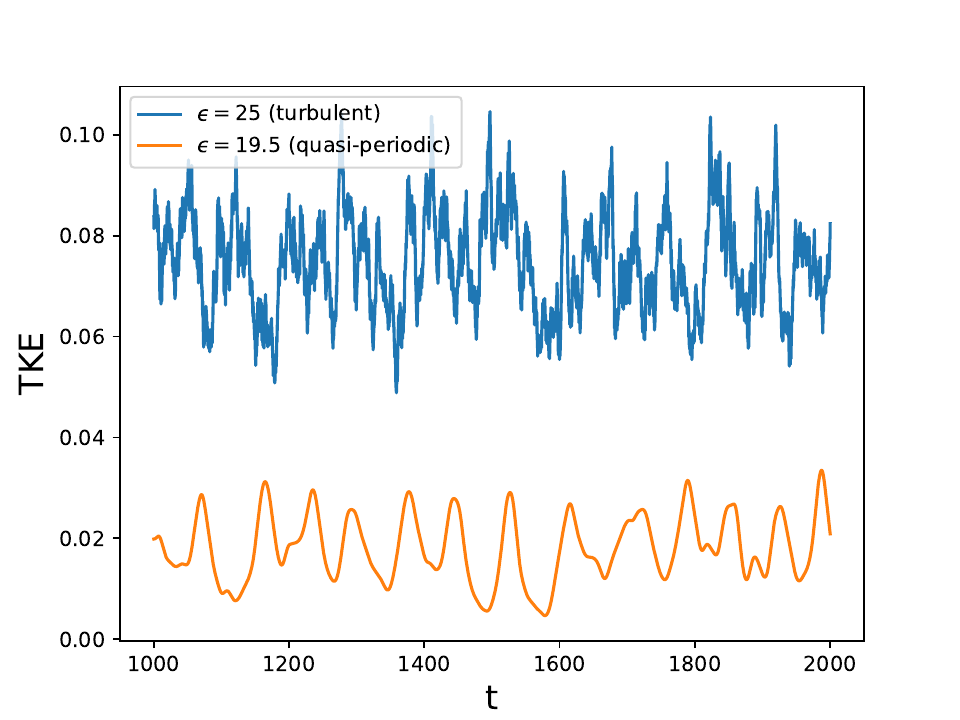}
\label{fig:1a}} 
\end{subfigure}
\begin{subfigure}{0.8\textwidth}\centering{ \caption{}
\includegraphics[width=0.75\linewidth]{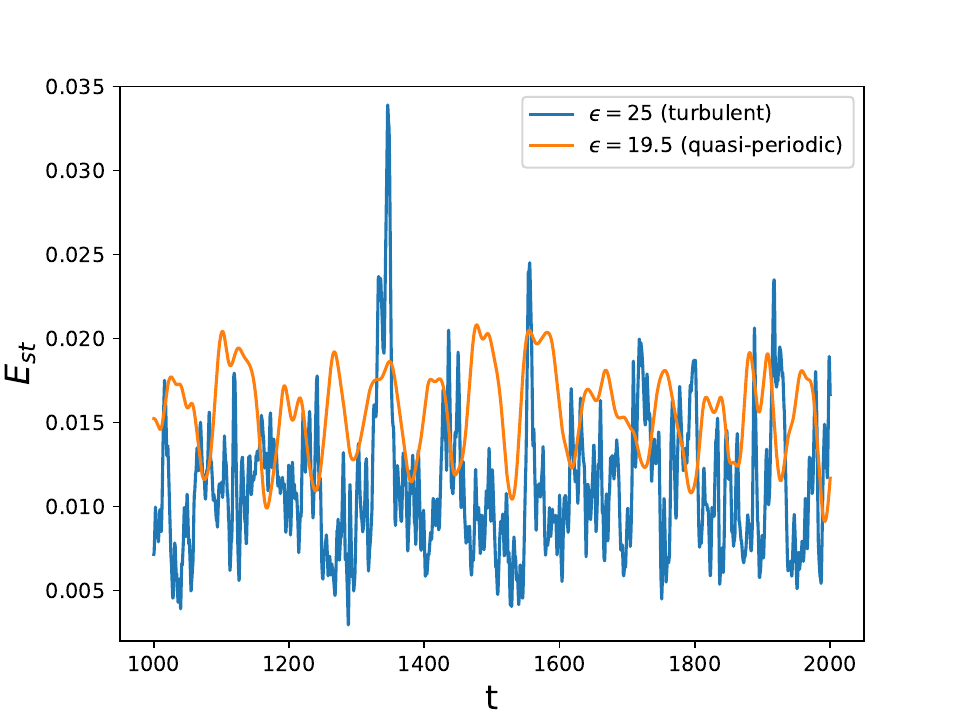} \label{fig:1b}} 
\end{subfigure}
\begin{subfigure}{0.8\textwidth} 
\centering{
\caption{}
\includegraphics[width=0.75\linewidth]{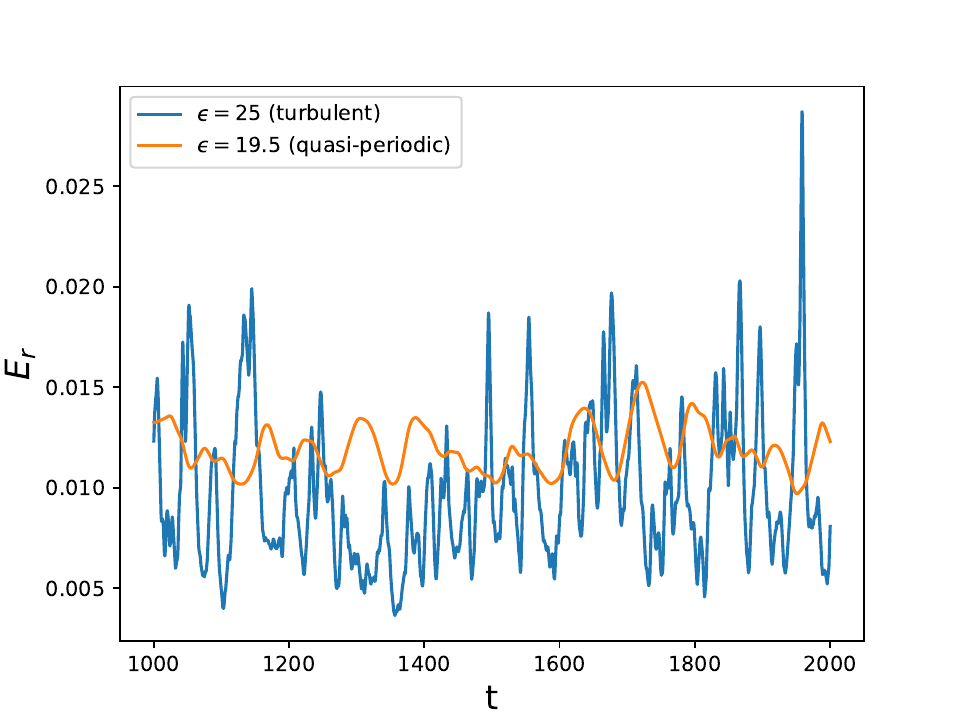} \label{fig:1c}} 
\end{subfigure}
}
\caption{Time series of perturbation, streak, and roll energy for quasi-periodic and turbulent equilibria. Perturbation $TKE$ is shown in (a), Streak energy, 
$E_{st}$, is shown in (b), and roll energy, $E_{r}$, is shown in (c).  Parameters for the limit cycle equilibrium are [$\epsilon =19.5$, $S=0.1$].  Parameters for the turbulent equilibrium are [$\epsilon =25$, $S=0.1$]; $Re=300$.}  
	\label{fig:flow}
    \end{figure}

Fixed-point S3T SSD equilibria are time-independent in streak energy, $E_{st}$, roll vorticity, $\Omega_r$, and TKE while both time-dependent and turbulent equilibria are time-dependent in these diagnostics.  Shown in figure \ref{fig:flow} are time series of $TKE$, $E_{st}$, and $E_{r}$ for an example of a time-dependent and of a turbulent equilibrium.

As the growth rate of the RSS instability increases with increasing S and/or $\epsilon$, the finite-amplitude equilibrium statistical state of the RSS proceeding from the instability becomes time-dependent rather than a fixed point. These time-dependent equilibria manifest either as approximate limit-cycle oscillations with nearly fixed period or, with further increase of instability, as quasi-periodic states.

 \begin{figure}
\subfloat[]{%
            \includegraphics[width=.48\linewidth]{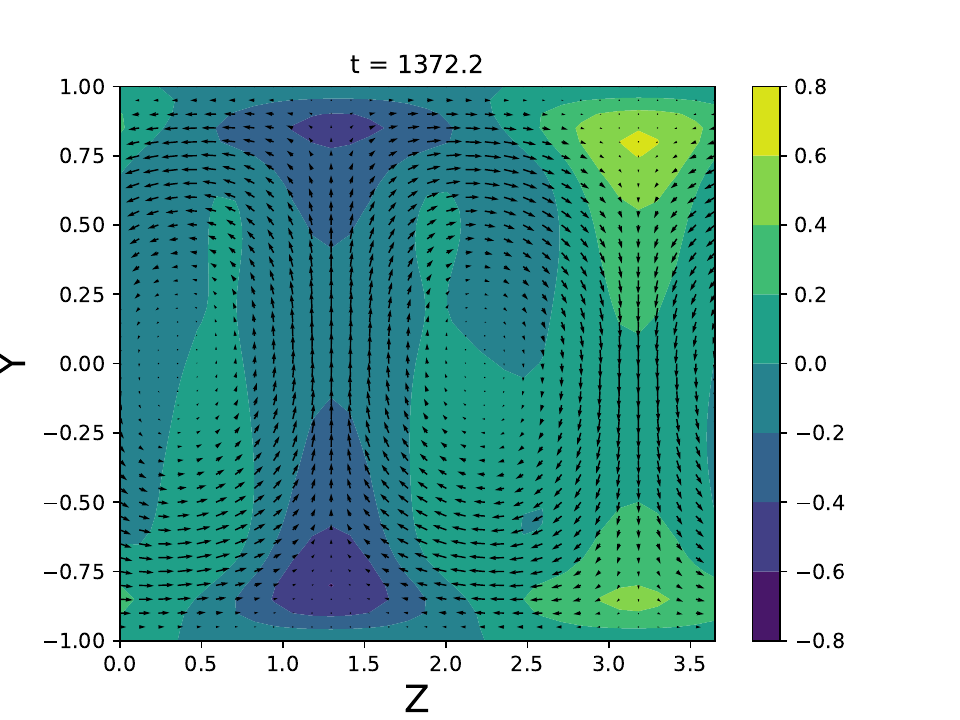}
            \label{subfig:a}%
        }\hfill
        \subfloat[]{%
            \includegraphics[width=.48\linewidth]{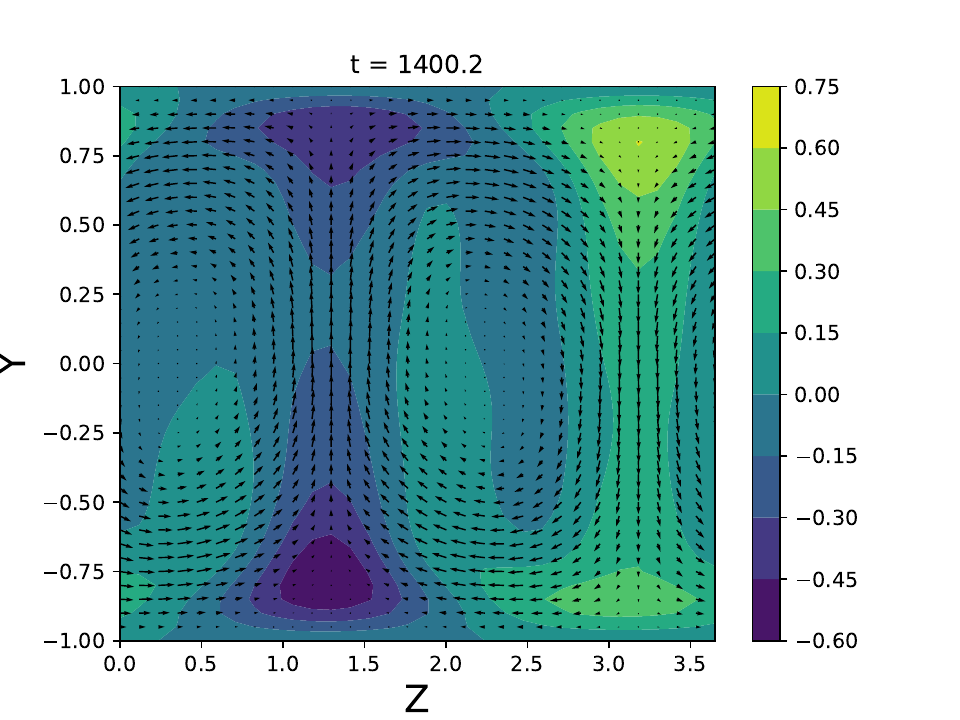}
            \label{subfig:b}%
        }\\
        \subfloat[]{%
            \includegraphics[width=.48\linewidth]{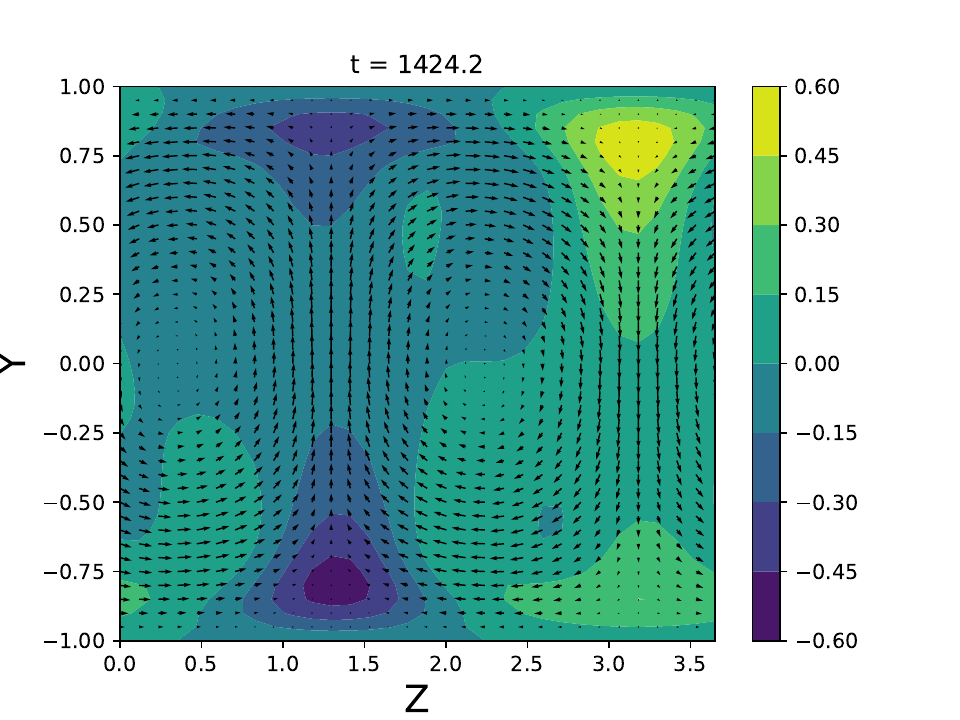}
            \label{subfig:c}%
        }\hfill
        \subfloat[]{%
            \includegraphics[width=.48\linewidth]{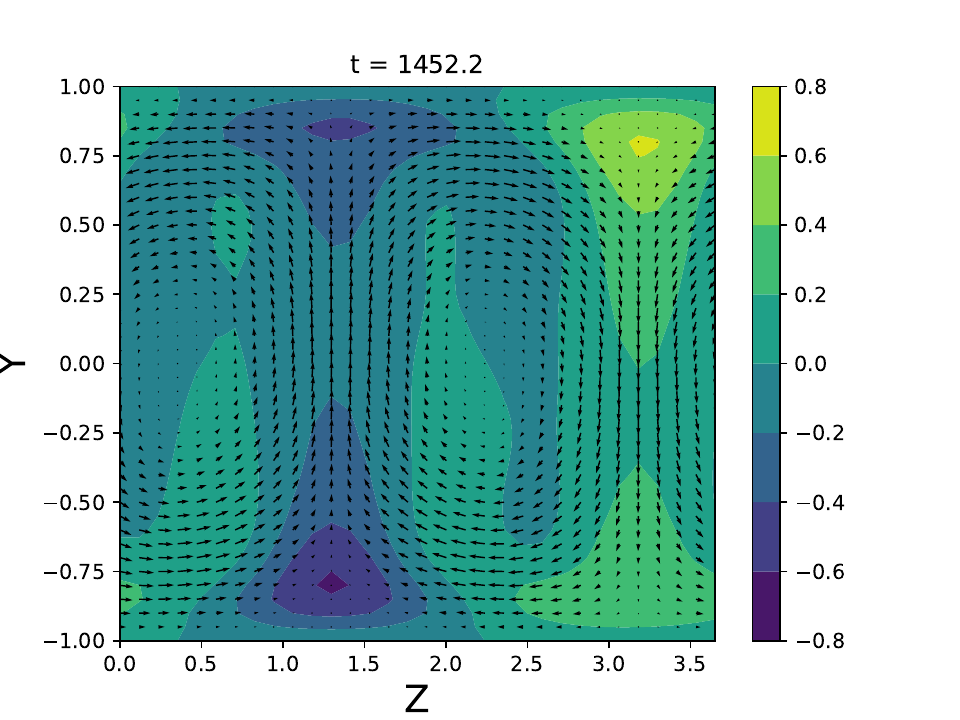}
            \label{subfig:d}%
        }
      
\caption{Snapshots of quasi-periodic time-dependent equilibria. Parameters are [$\epsilon =19.5$, $S=0.1$]; $Re=300$.}
\label{fig:timedependentsnapshots}
\end{figure}

\begin{figure}
\subfloat[]{%
            \includegraphics[width=.48\linewidth]{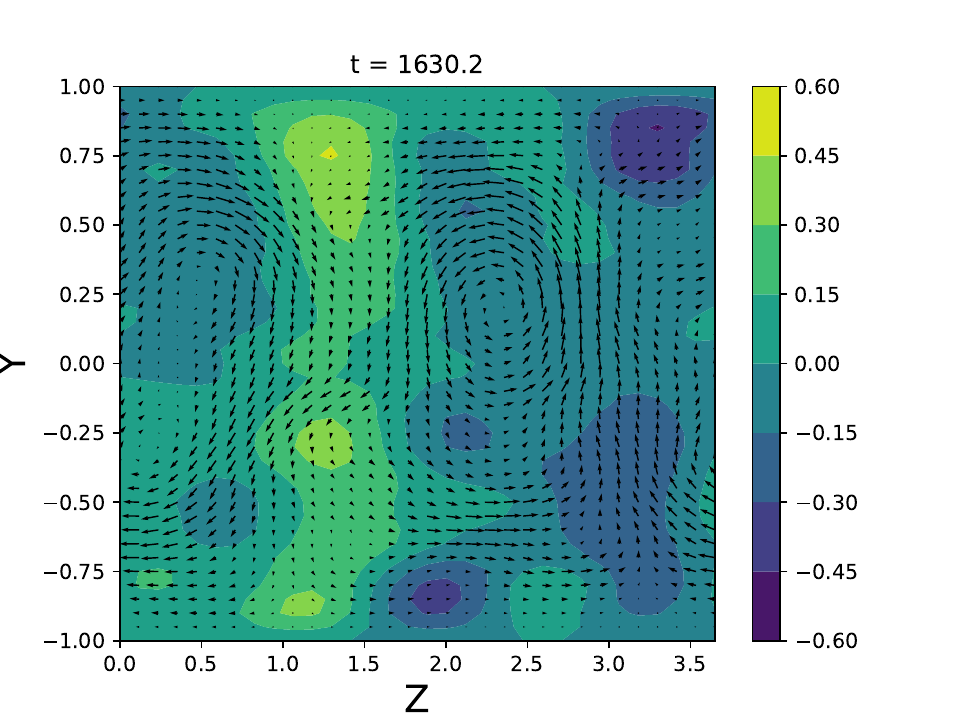}
            \label{subfig:a}%
        }\hfill
        \subfloat[]{%
            \includegraphics[width=.48\linewidth]{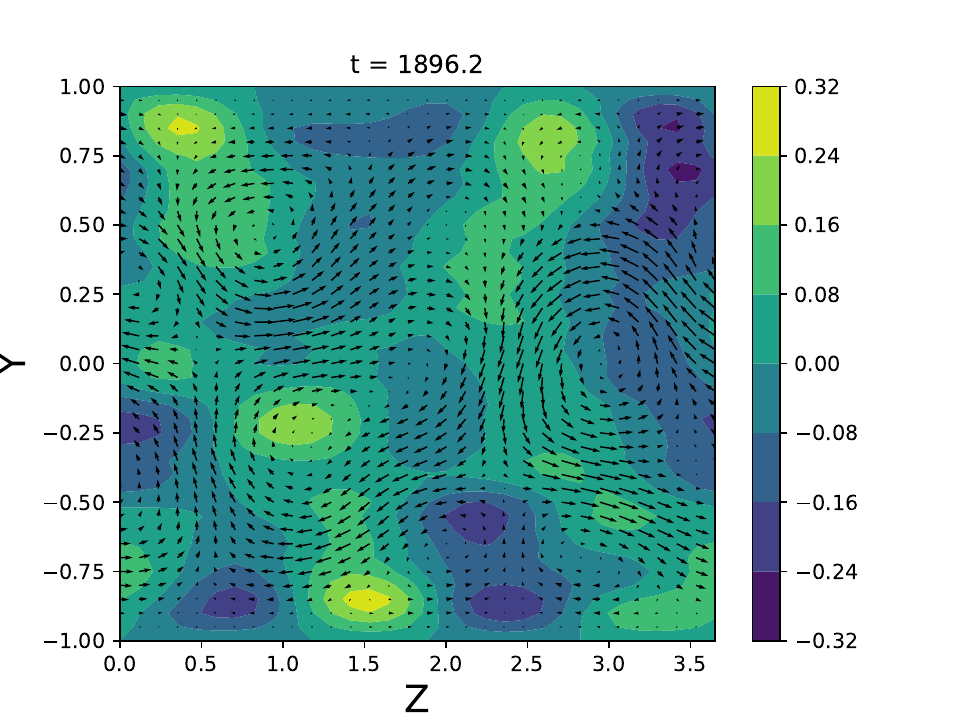}
            \label{subfig:b}%
        }\\
        \subfloat[]{%
            \includegraphics[width=.48\linewidth]{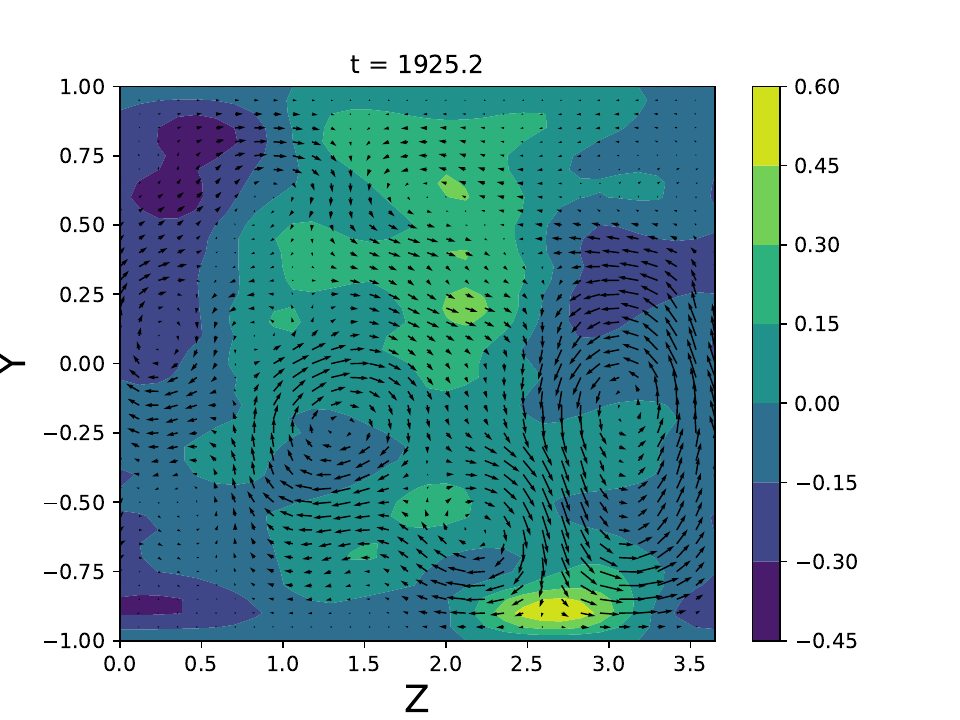}
            \label{subfig:c}%
        }\hfill
        \subfloat[]{%
            \includegraphics[width=.48\linewidth]{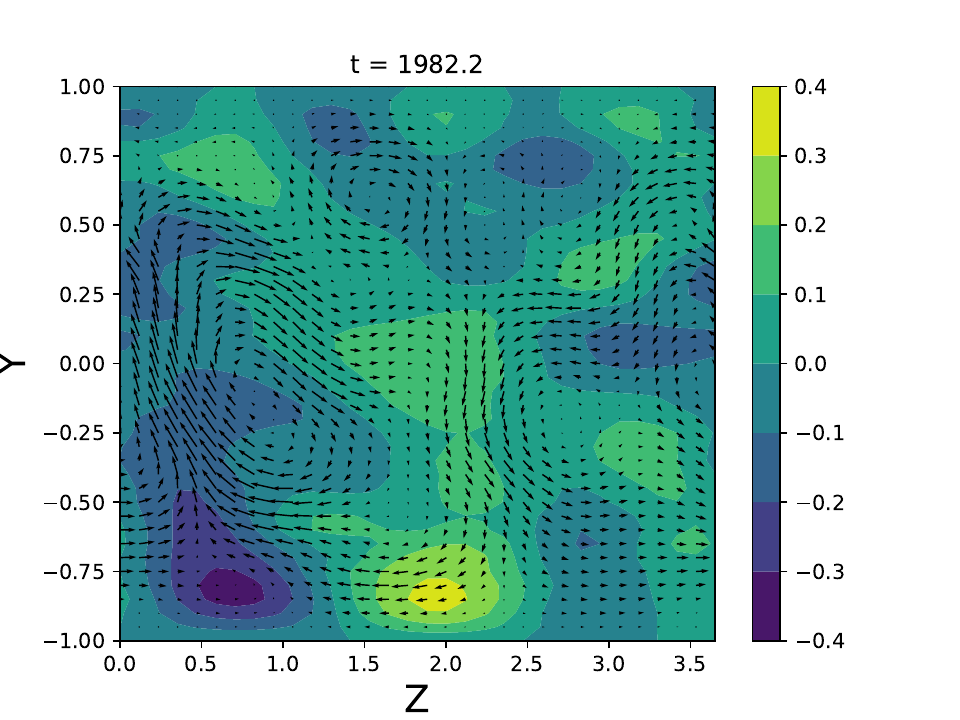}
            \label{subfig:d}%
        }
      
\caption{Snapshots of turbulent equilibria.   Parameters are [$\epsilon =25$, $S=0.1$]; $Re=300$. }
\label{fig:turbulentsnapshots}
\end{figure}

Representative snapshots of time-dependent equilibria are shown in Figure \ref{fig:timedependentsnapshots}. As the instability parameters (S and/or $\epsilon$) increase further, the system eventually transitions to turbulence.  Snapshots of turbulent equilibria are shown in Figure \ref{fig:turbulentsnapshots}.

It is important to note that the time dependence seen in these examples is intrinsic to the S3T SSD non-linear dynamics and that no time-dependent excitation is being imposed in these simulations.  It is also noteworthy that there is a robust nonlinear S3T mode underlying the dynamics that dominates the solution state at low super-criticality resulting in a clear quasi-periodic cycle while at higher super-criticality self-advection of the rolls produces the chaotic solution, which we refer to as turbulent here,  while retaining an underlying quasi-periodicity.

\section{RSS maintenance mechanism}
We now consider the mechanism maintaining the RSS in CL dynamics and its relation to the mechanism maintaining the RSS in wall-bounded shear flows \citep{Farrell 2016}.
The mechanism underlying RSS dynamics and equilibration can be examined using the balance equations for the RSS roll and streak components $\Omega_{r}$, and $U_{st}$ \citep{Farrell 2016}.  The equation governing the balance of the streak component is:\\
\begin{equation}
\partial_t U_{st}=-(\partial_y(UV)-\partial_y[UV]_z)-\partial_z(UW)-(\partial_y[u'v']_x-\partial_y[u'v']_{x,z})-\partial_z[u'w']_x+\Delta_1 \frac{U_{st}}{Re}-\frac{W-[W]_z}{\Gamma},
\end{equation}\\

Given that streaks of both signs occur, in order to obtain a linear measure of streak forcing, each term is multiplied by $sign(U_{st})$.  The equations for the physically distinct terms in the dynamics maintaining $U_{st}$ are:\\

\begin{equation}I_A=sign(U_{st})\cdot(-(V\frac{\partial U}{\partial y}-[V\frac{\partial U}{\partial y}]_z)-(W\frac{\partial U}{\partial z}-[W\frac{\partial U}{\partial z}]_z))\end{equation}
\begin{equation}
I_B=sign(U_{st})\cdot(-([v'\frac{\partial u'}{\partial y}]_x-[v'\frac{\partial u'}{\partial y}]_{x,z})-([w'\frac{\partial u'}{\partial z}]_x-[w'\frac{\partial u'}{\partial z}]_{x,z}))
\end{equation}
\begin{equation}
I_C=sign(U_{st})\cdot(\frac{1}{Re}\Delta_1 U_{st})
\end{equation}\\

These components of streak forcing are identified as lift-up ($I_A$), Reynolds stress ($I_B$), and viscous damping, ($I_C$). 

The balance equation for streamwise mean vorticity, which is indicative of roll maintenance, is:\\
\begin{equation}
\partial_t\Omega_r=-(V\partial_y +W \partial_z)\Omega_r+(\partial_{zz}-\partial_{yy})(\overline{v'w'})-\partial_{yz}(\overline{w'^2}-\overline{v'^2})+\Delta_1\frac{\Omega_r}{Re}+S\frac{\partial U_{st}}{\partial z}
\end{equation}

\begin{equation}
I_D=sign(\Omega_r) \cdot (-(V\partial_y +W \partial_z)\Omega_r)
\end{equation}
\begin{equation}
I_F=sign(\Omega_r) \cdot ((\partial_{zz}-\partial_{yy})(\overline{v'w'})-\partial_{yz}(\overline{w'^2}-\overline{v'^2}))
\end{equation}
\begin{equation}
I_G= sign(\Omega_r) \cdot \Delta_1\frac{\Omega_r}{Re}
\end{equation}
\begin{equation}
I_H=sign(\Omega_r) \cdot S\frac{\partial U_{st}}{\partial z}
\end{equation}

\begin{figure}
\centering{
\begin{subfigure}{0.8\textwidth} \caption{}
\includegraphics[width=.90\linewidth]
{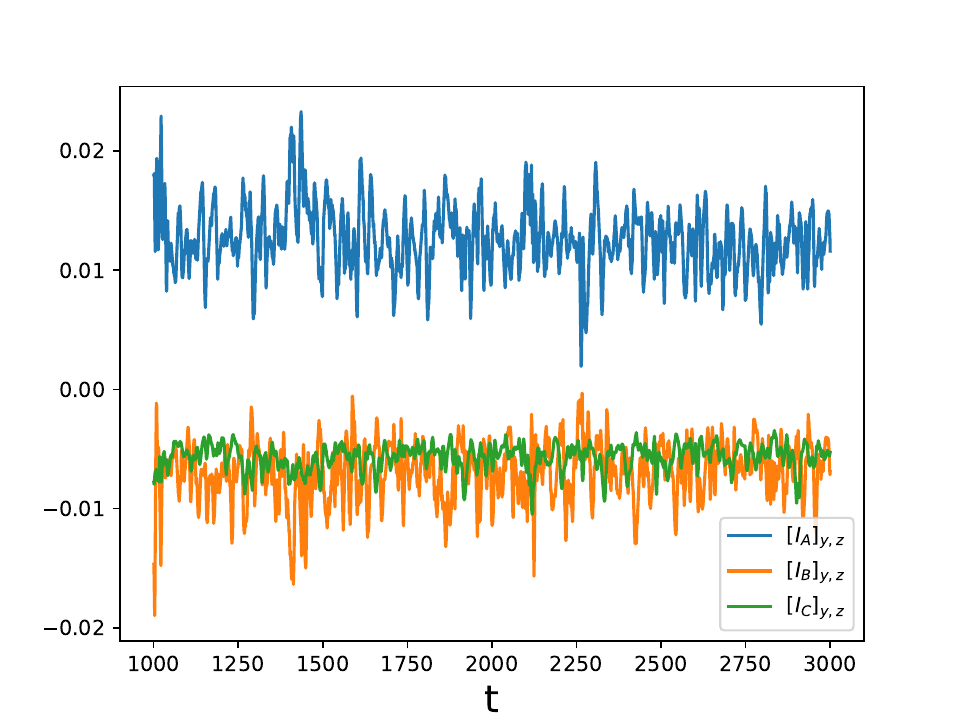}
\end{subfigure}
            \begin{subfigure}{0.8\textwidth} \caption{}

            \includegraphics[width=.90\linewidth]{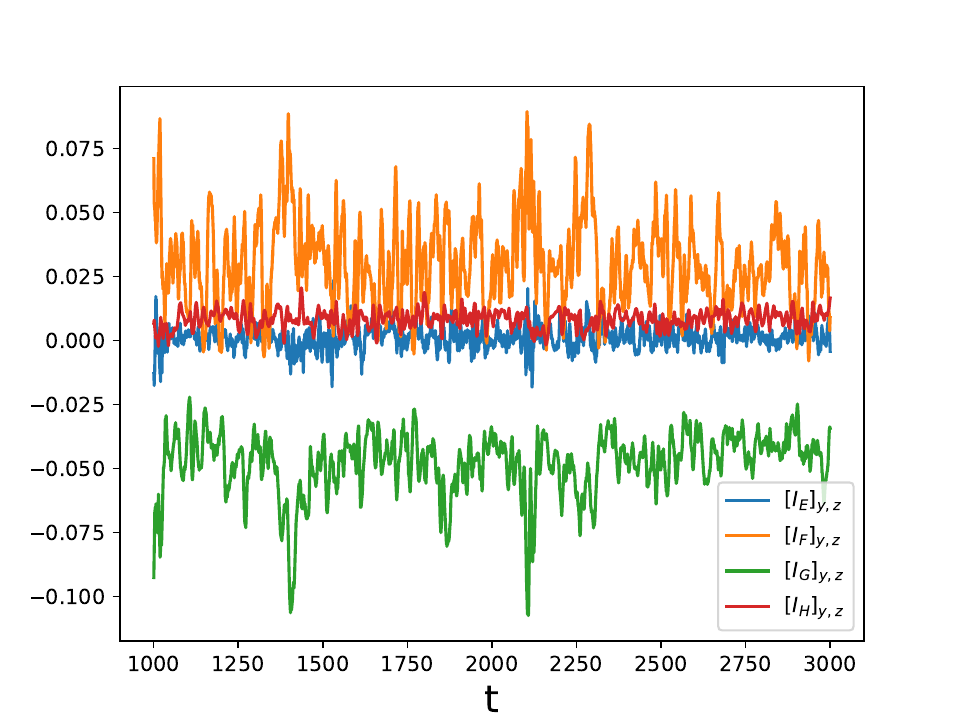}
            \end{subfigure}
 }       
      
\caption{Components of the balance maintaining the streak and roll in the  turbulent equilibrium for parameters [$\epsilon =25$, $S=0.1$]; $Re=300$; 
(a) streak maintenance and (b) roll maintenance.  }
\label{fig:roll/streakmaintenance}
\end{figure}
These components of roll forcing are identified as advection, ($I_D$), Reynolds stress, ($I_F$), viscous damping, ($I_G$), and Stokes drift tilt, ($I_H$).

Time series of the components maintaining the streak and roll in Langmuir RSS turbulence are shown in Figure \ref{fig:roll/streakmaintenance}.  The streak maintenance mechanism is similar to that in wall-bounded shear flow turbulence in which lift-up balances transfer of energy from the streak by the resolved perturbation Reynolds stresses, which are maintaining the perturbation TKE by extracting energy from the streak, and viscous dissipation. Maintenance of the roll vorticity is also similar to that in wall-bounded shear flow turbulence in which the Reynolds stress term balances dissipation except for a minor additional vorticity source from the Stokes drift tilt occurring in the Langmuir turbulence case.

\section{Self-sustaining Langmuir Turbulence}
The turbulent state supported by the S3T SSD implementation of CL dynamics is self-sustaining in the sense that it persists if the sochastic excitation parameterization is removed by setting $\epsilon=0$. When $\epsilon=0$, the S3T equation reduces to 
\begin{equation}
\frac{d \Gamma}{dt}=G(\Gamma) + \sum_kL_{RS}C_k
\end{equation}
\begin{equation}
\frac{d C_k}{dt}=A C_k+C_k A^{\dagger}
\end{equation}
We find that all turbulent equilibria in our simulations are self-sustaining.
This result is consistent with our analysis of the maintenance of the streak $U_{st}$ and roll $\Omega_r$, which are indicative of the underlying mechanism sustaining Langmuir RSS turbulence being the same self-sustaining process (SSP) familiar in the context of the dynamics of wall-bounded shear flow turbulence, apart from the minor positive roll vorticity forcing arising from the Stokes drift tilt.

\begin{figure}
\centering{
\begin{subfigure}{0.75\textwidth} \centering{ \caption{}
\includegraphics[width=.75\linewidth]{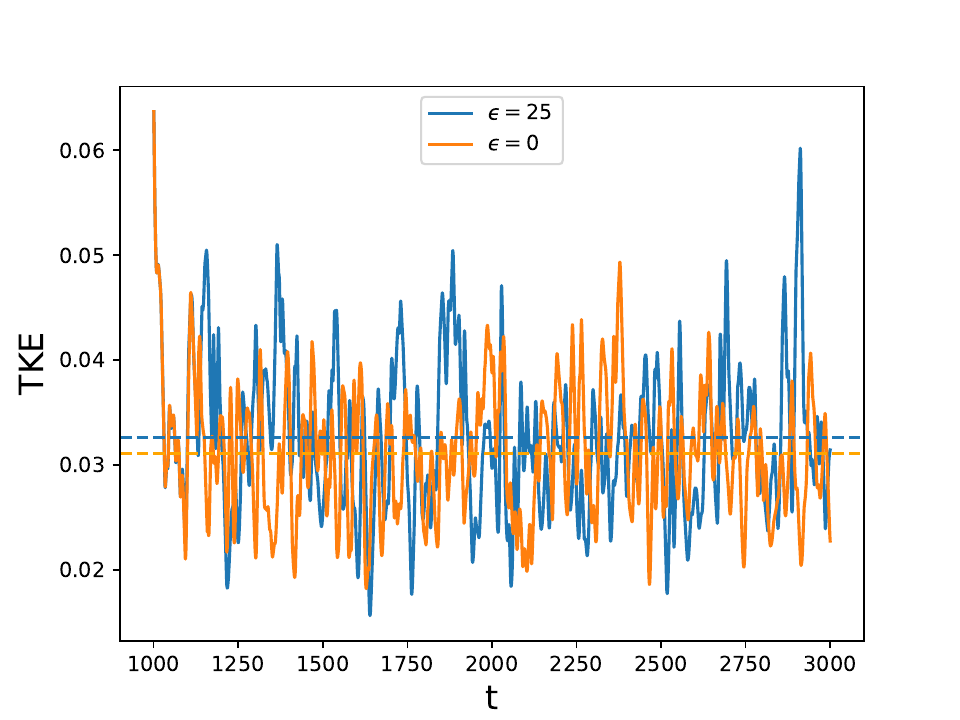}}
\end{subfigure}
\begin{subfigure}{0.75\textwidth} \centering{\caption{}
\includegraphics[width=.75\linewidth]{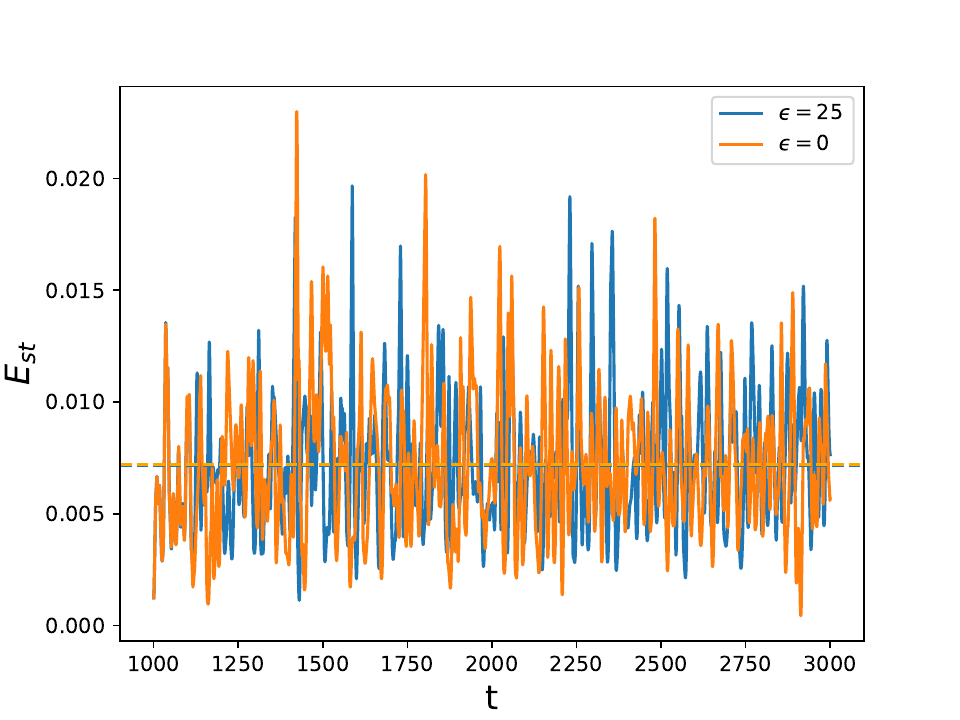}}
\end{subfigure}
\begin{subfigure}{0.75\textwidth} \centering{\caption{}
\includegraphics[width=.75\linewidth]{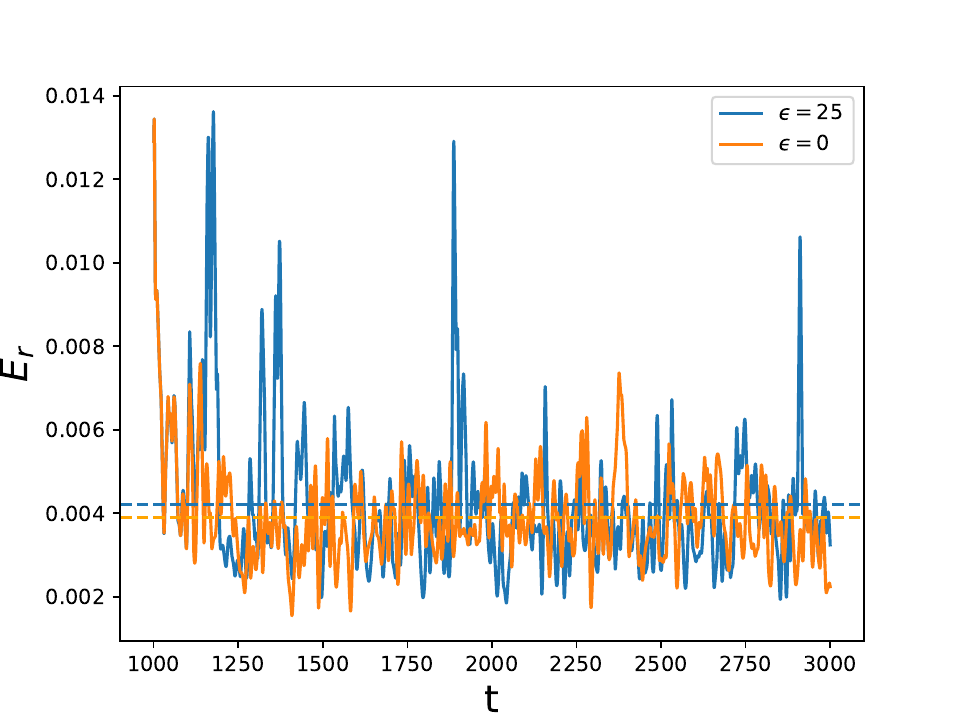}}
\end{subfigure}
}
\caption{Turbulence diagnostics for the turbulence with the background turbulence excitation that was used to instigate transition to the turbulent state retained ($\epsilon=25$) (blue) and the self-sustained turbulence maintained after the instigating turbulence excitation has been removed, $\epsilon=0$ (orange).  Time series of $TKE$ is shown in (a), of  
streak energy, $E_{st}$, in (b), and of roll energy, $E_{r}$, in (c).   Dashed blue lines indicate time averaged $TKE$ and $E_{st}$ for the case with $\epsilon=25$ retained. Similarly, dashed orange lines indicate time averaged $TKE$ and $E_{st}$ for the self-sustained case with $\epsilon=0$. Additional parameters  $S=0.1$, $Re=300$}
\label{fig:tkeEsssp}
\end{figure}
For example, the turbulent state at $S=0.1$ and $\epsilon=25$ initiated by the unstable S3T mode at these parameters equilibrates to a turbulent state.   Upon removing the stochastic forcing by setting $\epsilon=0$, this tubulent state persists as verified by comparison between time series diagnostics of these turbulent states shown in Figure \ref{fig:tkeEsssp}. 

While the  streak energy, $E_{st}$, and the perturbation TKE are greater when
$\epsilon=25$, both are sustained primarily by the same SSP as that sustaining wall-bounded shear flow turbulence. Greater TKE is expected in the case with parameterized stochastic forcing as $\epsilon$ directly injects energy into the perturbation covariance $\sum_k C_k$.  The increase in the roll and streak energy results from this exogenous turbulence positively contributing to the SSP by driving the roll directly by Reynolds stress torque and the streak indirectly through the roll-induced lift-up process.

\section{Langmuir circulation instability and equilibration with surface-concentrated Stokes drift}
\begin{figure}
\centering{
\begin{subfigure}{0.55\textwidth} \caption{}
\includegraphics[width=\linewidth]{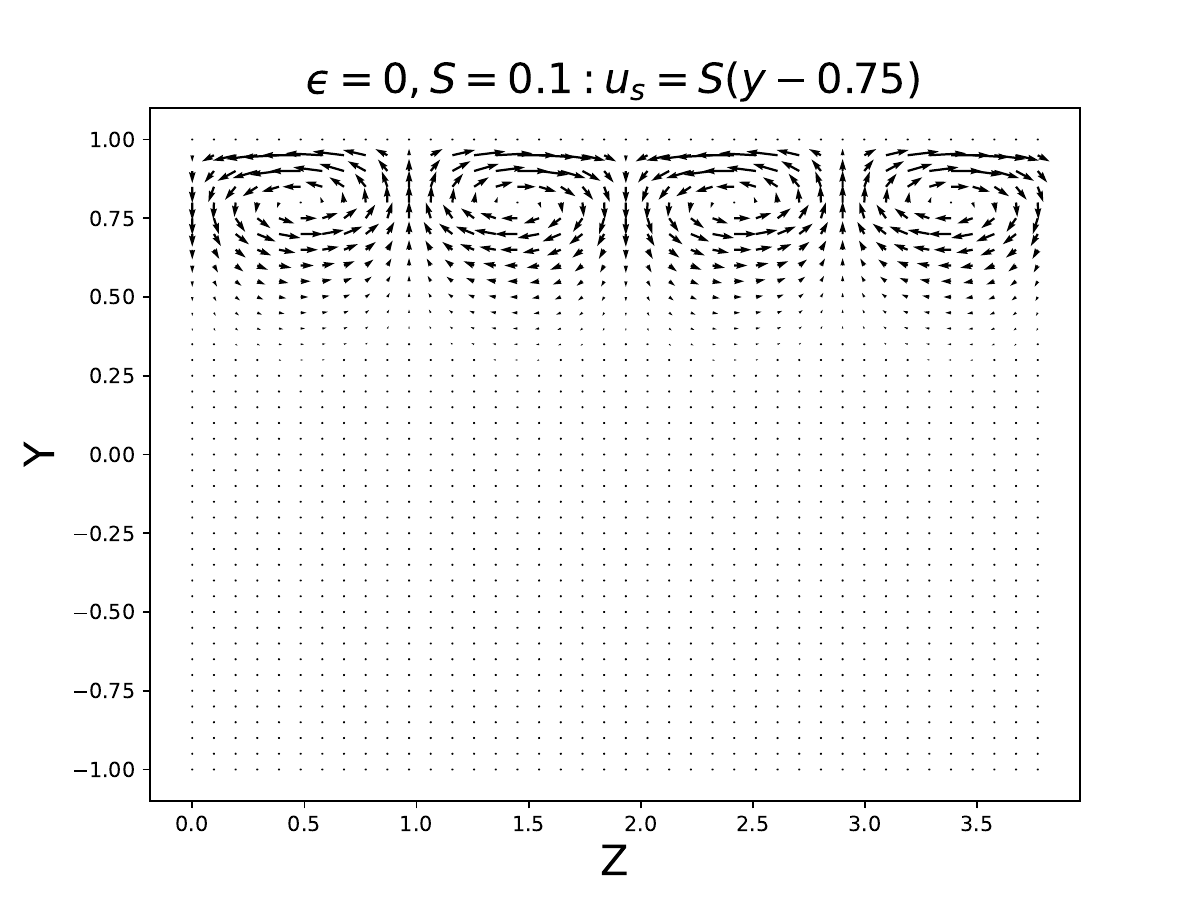}
\end{subfigure}
\begin{subfigure}{0.55\textwidth} \caption{}
\includegraphics[width=\linewidth]{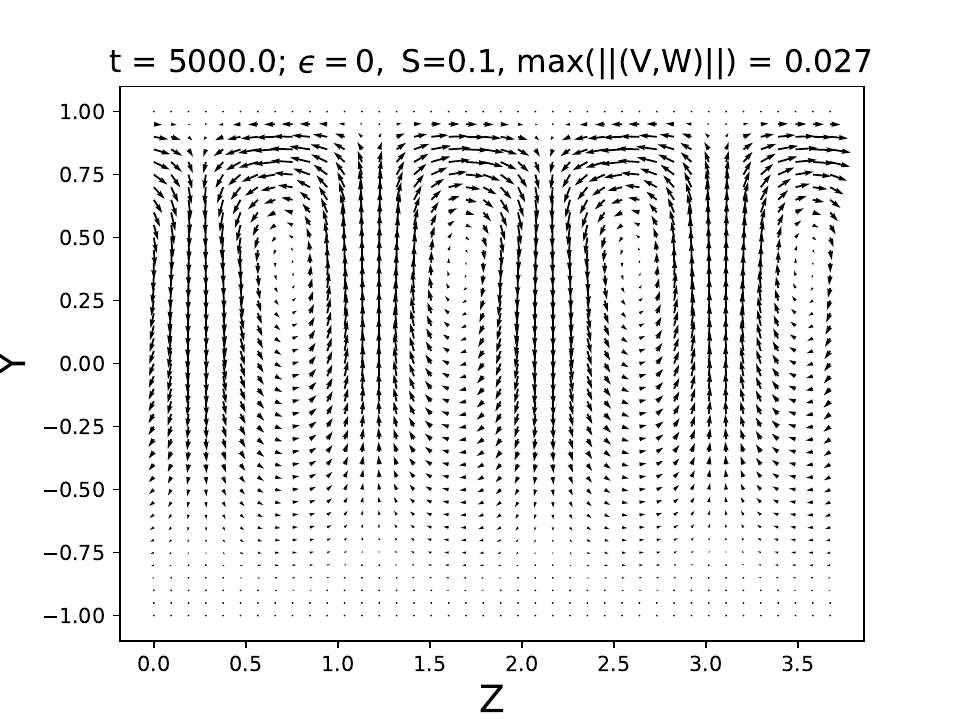}
\end{subfigure}
\begin{subfigure}{0.55\textwidth} \caption{}
\includegraphics[width=\linewidth]{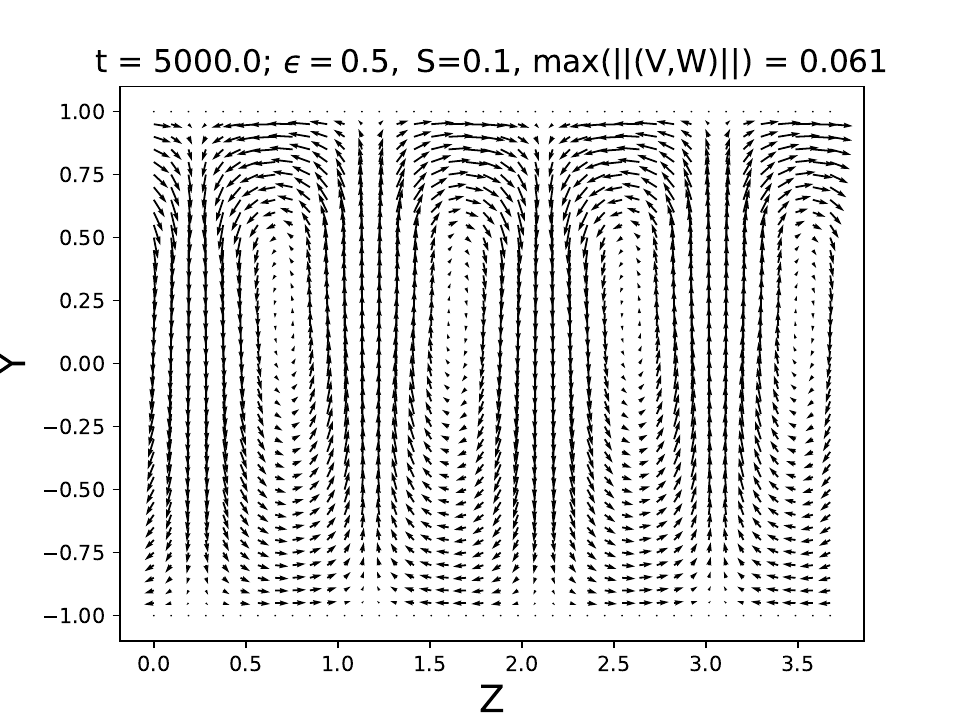}
\end{subfigure}

}
\caption{Examples showing that both nonlinear equilibration and the RS instability mechanism contribute to extending the pure CL2 eigenmode beyond the region of its Stokes drift support.  In (a) is shown the Langmuir CL2 instability mode for Stokes drift confined to the upper eighth of the mixed layer, (parameters $S=0.1,\epsilon=0$). The mode roll component, $(\delta V, \delta W)$, is shown by vectors.  In $(b)$  is shown the nonlinear equilibration of this eigenmode. In $(c)$ is shown the nonlinear equilibrium resulting from the combined CL2–RS torque mechanisms, (parameters $S=0.1,\epsilon=0.5$); $Re=300$.} 
\label{compareRSCLfortopquarterstokes}
\end{figure}

In previous sections, we assumed a distribution structure for Stokes drift velocity, $u_s(y)=S  \cdot y$, identical to the constant shear profile assumed for the Eulerian velocity. This simple choice allowed us to gain an understanding of how the CL2 instability mechanism combines with the Reynolds stress instability mechanism to form the Langmuir RSS. We also gained understanding using this simple model of the dynamics of non-linear equilibration of the RSS proceeding from these instabilities while retaining only adjustable parameters that are fundamentally related to the Langmuir circulation dynamics.  However, realistic Stokes velocity profiles are concentrated close to the top of the mixed layer, which motivates taking into account this dynamically important restriction in the Langmuir CL2 dynamics when it is being applied in a realistic physical context. In this section, we study the CL S3T SSD under restriction to a surface-confined Stokes velocity profile defined as: $u_s(y)=S\cdot (y-0.75)$ for $y\geq 0.75$ and $u_s(y)=0$ for $y<0.75$.\\
In order to study the Langmuir RSS stability  under this restriction to the Stokes velocity profile, we employ \eqref{meanS3Tlinear} and \eqref{covarianceS3Tlinear} repeating the stability analysis except using 
$u_s(y)=S\cdot (y-0.75)$ for $y\geq0.75$ and 
$u_s(y)=0$ for $y\leq0.75$.  
In Figure \ref{compareRSCLfortopquarterstokes} is shown the RSS eigenmode of the CL instability for parameters ($\epsilon=0,S=0.1$) with this surface-confined Stokes velocity profile.  Also shown is the equilibrium fixed-point RSS proceeding from this instability.  This result verifies the mechanism of downward extension of the surface-confined CL2 instability by self-advection in the nonlinear equilibration process.  However, inclusion of background turbulence results in an equilibrium with a more robust downward extension, as shown in panel (c).

\section{Discussion and Conclusions}
The ubiquity of RSS in observations of shear flows,  its fundamental role in the mechanism maintaining turbulence, and its dominant contribution to mixing of momentum and tracers in boundary layers motivates the study of RSS formation, maintenance, and equilibration.  A core challenge to this study is the fact that, although RSS is a structure of optimal growth in laminar shear flow \citep{Butler-Farrell-1992}, it is not typically an unstable structure.  However, the RSS is easily destabilized by a mechanism transferring momentum from the streak to the roll component of the RSS, such as a small spanwise frame rotation.  In the case of the Langmuir CL2 instability, imbalance arising from the advection of cross-stream Eulerian shear vorticity by the Stokes drift shear lacking a compensating advection of spanwise Stokes vorticity by the Eulerian spanwise shear produces a roll perturbation forcing proportional to the Eulerian streak perturbation which destabilizes the Langmuir RSS.  In the case of wall-bounded shear flows, a streak perturbation strains the background field of turbulence, producing a distribution of RS that forces the roll proportional to the streak perturbation, which destabilizes the RSS.  In the ocean mixed layer, both mechanisms contribute synergistically to destabilize the RSS.

In our study of the Langmuir circulation problem, we formulated the CL dynamics in S3T SSD form.  The S3T SSD is a statistical state dynamics closed at second order in which the CL2  instability is contained in the first cumulant while the RS torque instability arises from the interaction between the first and second cumulants.  These instabilities can be examined independently and in synergy by varying the Stokes (S) and RS ($\epsilon$) parameters.
While the S3T SSD formulation is nonlinear, by using linear perturbation theory the eigenmodes of the pure CL2, pure RS torque, and synergistic combinations of these can be conveniently studied.  In addition, the same nonlinear S3T SSD contains the mechanism of equilibration of these eigenmodes.  We find that, as a function of parameters,  the S3T SSD RSS equilibrates to a fixed point, a quasiperiodic cycle, or to a turbulent state.  The turbulent state is found to be self-support in the sense that, once it has been established by bifurcation from an unstable S3T SSD equilibrium, the destabilizing background turbulence excitation parameter $\epsilon$ can be removed and the turbulence subsequently sustains itself through the self-sustaining process (SSP) familiar in the context of wall-bounded shear flows.

In the S3T SSD the dynamics of RSS equilibria is conveniently partitioned into physical mechanisms responsible for the statistical mean balance maintaining the streak, roll, and perturbation components.  In the case of turbulent RSS equilibria the dominant balance maintaining the streak is between lift-up by the roll and a combination of resolved RS interactions, which are maintaining the turbulent state by extracting energy from the streak, and viscous dissipation.  The dominant balance maintaining the roll is between RS torque and dissipation with the Stokes tilting term making a minor positive contribution.  

Given both observational and theoretical reasons to expect highly surface restricted regions satisfying conditions for CL2 instability, we used our S3T SSD to analyze the concept of ballistic downward extension of the CL2 eigenmode beyond the region of its Stokes drift support.  We found, in an example with Stokes drift confined to the upper eighth of the mixed layer, that the pure CL2 eigenmode is primarily confined to the region of its support by the Stokes drift, but that when the S3T SSD was run to equilibrium a substantial downward extension was seen. However, addition of modest support for the RS torque mechanism by parameterized background turbulence resulted in a much more robust penetration, extending all the way to the lower mixed layer boundary in our model.   

In summary, an analysis that incorporates the CL mechanism in the S3T SSD turbulence model provides a more comprehensive approach to understanding Langmuir circulations.  It predicts the synergistic interaction between the CL2 and RS torque destabilization mechanisms, existence of Langmuir turbulence, the self-sustaining process maintaining Langmuir turbulence, and substantial augmentation by the RS torque mechanism in the support of deep transport across the mixed layer.

\end{document}